\documentclass[submission,copyright,creativecommons]{eptcs}
\providecommand{\event}{ICLP 2025} 

\usepackage{iftex}

\ifpdf
  \usepackage{underscore}         
  \usepackage[T1]{fontenc}        
\else
  \usepackage{breakurl}           
\fi

\usepackage{alltt}
\usepackage{proof}
\usepackage{bussproofs}

\hypersetup{hidelinks}

\hypersetup{
  pdftitle={Implementing the First-Order Logic of Here and There},    
  pdfauthor={Jens Otten \& Torsten Schaub},
  pdfkeywords={Logic Programming, Here-and-There, Theorem Proving, leanHaT}
}

\newcommand{\leanhat}{\mbox{\sf leanHa$\!$T}}
\newcommand{\leanHAT}{\mbox{\sf leanHa\hspace{-0.7mm}T}}
\newcommand{\leancop}{\mbox{\sf leanCoP}}

\newcommand{\nanocop}{\mbox{\sf nanoCoP}}
\newcommand{\nanocopi}{\mbox{\sf nanoCoP-i}}
\newcommand{\leantap}{\mbox{{\sf lean}$T\!\!A\!P$}}
\newcommand{\ileantap}{\mbox{\sf ilean$\!$T$\!$AP}}
\newcommand{\ileansep}{\mbox{\sf ileanSeP}}
\newcommand{\nanocopihat}{\mbox{\sf nanoCoP-i-HT}}
\newcommand{\ileantaphat}{\mbox{\sf ilean$\!$T$\!$AP-HT}}
\newcommand{\ileanTAPhat}{\mbox{\sf ilean\hspace{-0.7mm}T\hspace{-0.7mm}AP-HT}}
\newcommand{\ileansephat}{\mbox{\sf ileanSeP-HT}}
\newcommand{\impl}{\rightarrow}
\newcommand{\equi}{\leftrightarrow}

\newcommand{\mor}{\vee}
\newcommand{\mand}{\wedge}
\newcommand{\ax}{\mbox{axiom}}
\newcommand{\anl}{\mbox{$\mand$-left}}
\newcommand{\anr}{\mbox{$\mand$-right}}
\newcommand{\orl}{\mbox{$\mor$-left}}
\newcommand{\orr}{\mbox{$\mor$-right}}
\newcommand{\iml}{\mbox{$\impl$-left}}
\newcommand{\imr}{\mbox{$\impl$-right}}
\newcommand{\nel}{\mbox{$\neg$-left}}
\newcommand{\ner}{\mbox{$\neg$-right}}
\newcommand{\all}{\mbox{$\forall$-left}}
\newcommand{\alr}{\mbox{$\forall$-right$^*$}}
\newcommand{\exl}{\mbox{$\exists$-left$^*$}}
\newcommand{\exr}{\mbox{$\exists$-right}}
\newcommand{\nall}{\mbox{$\neg\forall$-left$^*$}}
\newcommand{\nalr}{\mbox{$\neg\forall$-right}}
\newcommand{\nexl}{\mbox{$\neg\exists$-left}}
\newcommand{\nexr}{\mbox{$\neg\exists$-right$^*$}}
\newcommand{\negg}{$\neg$}
\newcommand{\HT}{\textbf{HT}}

\newcommand{\ep}{\varepsilon}
\newcommand{\fns}{\footnotesize}
\newcommand{\mfn}[1]{\mbox{\fns$#1$}}
\newcommand{\scr}{\scriptsize}
\newcommand{\bs}{\backslash}
\newcommand{\ul}[1]{\underline{#1}}
\newcommand{\vx}{\vec{x}}

\title{Implementing the First-Order Logic of Here and
 There}

\author{
Jens Otten
\institute{University of Pernambuco\\ Recife, Brazil}
\email{jeotten@leancop.de}
\and
Torsten Schaub 
\institute{University of Potsdam\\
Potsdam, Germany}
\email{torsten@cs.uni-potsdam.de}
}

\def\titlerunning{Implementing the First-Order Logic of Here
 and There}
\def\authorrunning{J. Otten \& T. Schaub}

\begin{document}
\maketitle

\newtheorem{definition}{Definition}
\newtheorem{example}{Example}

\begin{abstract}
We present automated theorem provers for the first-order logic of here and there (HT). They are based on a native sequent calculus for the logic of HT and an axiomatic embedding of the logic of HT into intuitionistic logic. The analytic proof search in the sequent calculus is optimized by using free variables and skolemization. The embedding is used in combination with sequent, tableau and connection calculi for intuitionistic first-order logic. All provers are evaluated on a large benchmark set of first-order formulas, providing a foundation for the development of more efficient HT provers.
\end{abstract}

\section{Introduction} 



The \emph{Logic of Here and There} (HT; \cite{heyting30a,goedel32a}) is an intermediate logic that is between classical logic (CL) and intuitionistic logic (IL).
HT is weaker than classical logic as --- just like intuitionistic logic --- it does not validate the law of excluded middle $A \vee \neg A$, giving it a more constructive character. However, it is stronger than intuitionistic logic as the so-called \emph{weak law of excluded middle} $\neg A \vee \neg\neg A$ is valid in HT.

HT plays a central role and provides the fundamental logical foundations for \emph{Answer Set Programming} (ASP; \cite{lifschitz19a}).
Specifically, the stable models of a logic program~\cite{gellif90a} can be characterized by a simple, minimal model construction within HT, and are termed equilibrium models~\cite{pearce06a}.
This characterization offers several advantages.
Notably, logical equivalence in HT implies substitutability without altering the equilibrium models of the containing formula~\cite{lipeva01a}. This property is crucial as it permits the simplification of a logic program segment independently of its context.
Furthermore, HT facilitates the characterization of ASP extensions incorporating foreign language constructs and inferences~\cite{agcadipevi13a,cakaossc16a,bedaei16a,cafafa20a}.
A first tableau system for HT and its extension for computing equilibrium models was presented in~\cite{peguva00a}.

While these works provide essential cornerstones for the advancement of ASP, they are propositional, contrasting with the first-order nature of ASP encodings.
To address this, first-order extensions of HT were introduced in~\cite{peaval05a}, and some of the above works lifted to the first-order case~\cite{lipeva07a}.
However, direct automated reasoning within HT remains undeveloped.
Current approaches, such as \emph{anthem}~\cite{falilusc20a,faglhaheliliscst25a}, which determines HT equivalence of two ASP encodings, rely on translations into Boolean first-order logic and employ conventional theorem provers for classical first-order logic.

We address this shortcoming by developing first-order theorem proving technology directly within HT, leading to four new theorem provers for HT. The first prover, \leanhat, is based on a first-order sequent calculus for HT. The other three provers, \ileansephat, \ileantaphat, \nanocopihat, are based on an axiomatic embedding of first-order HT into first-order intuitionistic logic and use the intuitionistic theorem provers \ileansep, \ileantap\ and \nanocopi, respectively. These intuitionistic provers are based on a single-succedent intuitionistic sequent calculus, a prefixed tableau calculus and a prefixed non-clausal connection calculus, respectively.

All four developed theorem provers for HT follow the \emph{lean methodology} for automated theorem provers as first propagated by the theorem provers \leantap~\cite{ar:BeckertPosegga95} and \leancop~\cite{ar:OttenBibel03} for classical first-order logic; they are implemented by using very compact yet efficient Prolog code.

The paper is structured as follows. Section 2 presents the sequent calculus LHT for HT and the axiomatic embedding of HT into intuitionistic logic. Section 3 describes proof search optimizations and methods. It includes techniques to prune the search space in the sequent calculus LHT, such as free variables and skolemization, and an optimized axiomatic embedding of HT into intuitionistic logic, which drastically reduces the number of axioms. We also provide details of the prefixed non-clausal connection calculus for intuitionistic logic, which is used by \nanocopihat\ and among the most efficient proof search calculi for intuitionistic logic. Section 4 presents implementation details of the four developed theorem provers for HT.
%
%
A comprehensive experimental evaluation of the theorem provers for HT on a large set of benchmark formulas is provided in Section 5, before the paper concludes with a summary and outlook in Section 6.

\section{Preliminaries} 

The following section describes syntax, semantics and a sequent calculus for the logic of HT, as well as an axiomatic embedding of 
the logic of HT into intuitionistic logic.

\subsection{Syntax and Semantics of HT}

The logic of HT has the same \emph{syntax} as classical and intuitionistic logic, but a different semantics.
Propositional formulas are defined inductively from atomic (propositional) formulas ($p, q$) and 
the connectives $\mand$ (conjunction), $\mor$ (disjunction), $\impl$ (implication) and $\neg$ (negation).
This definition is extended to first-order logic by adding first-order predicates, variables, terms and the quantifiers $\forall$ (universal) and $\exists$ (existential).
A \emph{term} is either a variable ($x, y, z$) or $f(t_1,\ldots,t_n)$, in which $f$ is an $n$-ary function symbol 
and $t_1,\ldots,t_n$ are terms (for $n\geq 0$).
A \emph{(first-order) formula} ($A,B,F,G,H$) is defined inductively and is either an atomic formula $P(t_1,\ldots,t_n)$ for an $n$-ary predicate symbol~$P$ and terms $t_1, \ldots, t_n$ (for $n\geq 0$), or a formula of the form $G\mand H$, $G\mor H$, $G\impl H$, $\neg G$, $\forall x\, G$ and $\exists x\, G$ for formulas $G$, $H$ and variable~$x$.
A literal $L$ has the form $A$ or $\neg A$ for an atomic formula $A$.
The \emph{free variables} in a formula are those not in the scope of a quantifier. $A[x\bs t]$ denotes the formula obtained by substituting the term $t$ for all free variable occurrences of $x$ in $A$. $\forall x_1,\ldots,x_n$ and $\exists x_1,\ldots,x_n$ are short for $\forall x_1 \ldots \forall x_n$ and $\exists x_1 \ldots \exists x_n$, respectively.

The \emph{semantics} of HT is defined by an intuitionistic Kripke-structure with two worlds $h$, $t$ (``here'' and ``there'') with constant individual domain and a reflexive and transitive accessibility relation between these two worlds. The connectives and quantifiers are interpreted classically in each single world except for implication and negation, which are interpreted with respect to both worlds. A formula that is true in $h$ is also true in $t$. See~\cite{lipeva07a} for more details.


\subsection{Sequent Calculus for HT}
\label{ssec:lht}

\def\fCenter{ ~\vdash~ }
\newcommand{\gCenter}{\vdash}

A \emph{sequent calculus} allows us to derive all \emph{valid} formulas of a given logic.
Sequent calculi for classical and intuitionistic first-order logic were first introduced by Gentzen~\cite{ar:Gentzen35}.
A formal (sequent) calculus consists of rules and axioms. Each rule has one or more premises and one conclusion, axioms have one conclusion but no premises.
For examples, the sequent calculus for classical logic consists of two rules for each connective and quantifier and one axiom.
The premises and conclusions are \emph{sequents} of the form
{$\Gamma \vdash \Delta$} where $\Gamma$ and $\Delta$ are multisets
of (first-order) formulas. $\Gamma,A$ or $A,\Gamma$ is short for 
$\Gamma\cup \{A\}$.
A \emph{sequent proof} of a (first-order) formula $F$ is a derivation tree of the sequent $\,\gCenter F$ (i.e.\ of $\{\}\gCenter \{F$\}) applying the rules of the sequent calculus, in which all leaves are axioms.

A sequent calculus for the first-order logic of HT was given by Mints~\cite{ar:Mints10}.
It consist of two axioms and two rules for each of the connectives and quantifiers $\mand$, $\mor$, $\impl$, $\forall $ and $\exists$. Additionally, there are altogether 12 rules for negation $\neg$ in combination with each of  the connectives and quantifiers.

\begin{definition}[Sequent Calculus LHT for HT]
The axioms and rules of the sequent calculus LHT for HT are given in Figure \ref{fig:seqhat}~\cite{ar:Mints10}.%
\footnote{
The {\negg\orr}-rule of the original calculus was slightly corrected by moving the formulas $\neg A$ and $\neg B$ from the left to the right side in the sequent of the premises. Otherwise, for example, the formula $(p \mor p) \impl p$, which is valid in HT, is not provable and the formula $\neg p \mor \neg (p \mor p)$, which is not valid in HT, is provable.} 
$t$ is an arbitrary term and for the rules marked with $\,^*$
the \emph{Eigenvariable condition} has to be respected:
the Eigenvariable $z$ must not occur in the conclusion, i.e.\  in $\Gamma$, $\Delta$, or $A$.
\end{definition}

Then, a formula $F$ is valid in HT if, and only if, there is a sequent proof for $F$ in LHT~\cite{ar:Mints10}.

\begin{figure}[hb!]

\hspace*{2.4mm}
\fbox{
\begin{minipage}{15.2cm}

\bigskip

\begin{tabular}{@{\hspace{1mm}}r@{\hspace{2mm}}r@{\hspace{1cm}}r@{\hspace{2mm}}r}


\AxiomC{}
\RightLabel{\ax$_1$}
\UnaryInfC{$\Gamma, A \fCenter A, \Delta$}
\DisplayProof
&
&
\AxiomC{}
\RightLabel{\ax$_2$}
\UnaryInfC{$\Gamma, A, \neg A \fCenter \Delta$}
\DisplayProof
\\[5mm]


\Axiom$\Gamma, A, B \fCenter \Delta$
\RightLabel{\anl}
\UnaryInf$\Gamma, A \mand B \fCenter \Delta$
\DisplayProof
&
{\em(r1)}
&
\Axiom$\Gamma \fCenter A, \Delta$
\Axiom$\Gamma \fCenter B, \Delta$
\RightLabel{\anr}
\BinaryInf$\Gamma \fCenter A\mand B, \Delta$
\DisplayProof
&
{\em(r8)}
\\[5mm]

\Axiom$\Gamma, A \fCenter \Delta$
\Axiom$\Gamma, B \fCenter \Delta$
\RightLabel{\orl}
\BinaryInf$\Gamma, A\mor B \fCenter \Delta$
\DisplayProof
&
{\em(r9)}
&
\Axiom$\Gamma \fCenter A, B, \Delta$
\RightLabel{\orr}
\UnaryInf$\Gamma \fCenter A \mor B, \Delta$
\DisplayProof
&
{\em(r2)}
\\[5mm]


\multicolumn{4}{l}{
\Axiom$\Gamma, \neg A \fCenter \Delta$
\Axiom$\Gamma \fCenter A, \neg B, \Delta$
\Axiom$\Gamma, B \fCenter \Delta$
\RightLabel{\iml}
\TrinaryInf$\Gamma, A \impl B \fCenter \Delta$
\DisplayProof
~\,{\em(r14)}
}
\\[5mm]

\multicolumn{3}{r}{
\Axiom$\Gamma, A \fCenter B, \Delta$
\Axiom$\Gamma, \neg B \fCenter \neg A, \Delta$
\RightLabel{\imr}
\BinaryInf$\Gamma \fCenter A \impl B, \Delta$
\DisplayProof
}
&
{\em(r13)}
\\[5mm]


\Axiom$\Gamma, \neg A \fCenter \Delta$
\Axiom$\Gamma, \neg B \fCenter \Delta$
\RightLabel{\negg\anl}
\BinaryInf$\Gamma, \neg (A \mand B) \fCenter \Delta$
\DisplayProof
&
{\em(r10)}
&
\Axiom$\Gamma \fCenter \neg A, \neg B, \Delta$
\RightLabel{\negg\anr}
\UnaryInf$\Gamma \fCenter \neg (A \mand B), \Delta$
\DisplayProof
&
{\em(r3)}
\\[5mm]

\Axiom$\Gamma, \neg A, \neg B \fCenter \Delta$
\RightLabel{\negg\orl}
\UnaryInf$\Gamma, \neg (A \mor B) \fCenter \Delta$
\DisplayProof
&
{\em(r4)}
&
\Axiom$\Gamma \fCenter \neg A, \Delta$
\Axiom$\Gamma \fCenter \neg B, \Delta$
\RightLabel{\negg\orr}
\BinaryInf$\Gamma \fCenter \neg (A \mor B), \Delta$
\DisplayProof
&
{\em(r11)}
\\[5mm]

\Axiom$\Gamma, \neg B \fCenter \neg A, \Delta$
\RightLabel{\negg\iml}
\UnaryInf$\Gamma, \neg (A \impl B) \fCenter \Delta$
\DisplayProof
&
{\em(r5)}
&
\Axiom$\Gamma, \neg A \fCenter \Delta$
\Axiom$\Gamma \fCenter \neg B, \Delta$
\RightLabel{\negg\imr}
\BinaryInf$\Gamma \fCenter \neg (A \impl B), \Delta$
\DisplayProof
&
{\em(r12)}
\\[5mm]

\Axiom$\Gamma \fCenter \neg A, \Delta$
\RightLabel{\negg\nel}
\UnaryInf$\Gamma, \neg\neg A \fCenter \Delta$
\DisplayProof
&
{\em(r6)}
&
\Axiom$\Gamma, \neg A \fCenter \Delta$
\RightLabel{\negg\ner}
\UnaryInf$\Gamma \fCenter \neg\neg A, \Delta$
\DisplayProof
&
{\em(r7)}
\\[5mm]


\Axiom$\Gamma, A[x\bs t], \forall x\, A \fCenter \Delta$
\RightLabel{\all}
\UnaryInf$\Gamma, \forall x\, A \fCenter \Delta$
\DisplayProof
&
{\em(r25)}
&
\Axiom$\Gamma \fCenter A[x\bs z], \Delta$
\RightLabel{\alr}
\UnaryInf$\Gamma \fCenter \forall x\, A, \Delta$
\DisplayProof
&
{\em(r21)}
\\[5mm]


\Axiom$\Gamma, A[x\bs z] \fCenter \Delta$
\RightLabel{\exl}
\UnaryInf$\Gamma, \exists x\, A \fCenter \Delta$
\DisplayProof
&
{\em(r22)}
&
\Axiom$\Gamma \fCenter \exists x\, A, A[x\bs t], \Delta$
\RightLabel{\exr}
\UnaryInf$\Gamma \fCenter \exists x\, A, \Delta$
\DisplayProof
&
{\em(r26)}
\\[5mm]


\Axiom$\Gamma, \neg A[x\bs z] \fCenter \Delta$
\RightLabel{\nall}
\UnaryInf$\Gamma, \neg (\forall x\, A) \fCenter \Delta$
\DisplayProof
&
{\em(r19)}
&
\Axiom$\Gamma \fCenter \neg (\forall x\, A), \neg A[x\bs t], \Delta$
\RightLabel{\nalr}
\UnaryInf$\Gamma \fCenter \neg (\forall x\, A), \Delta$
\DisplayProof
&
{\em(r23)}
\\[5mm]


\Axiom$\Gamma, \neg A[x\bs t], \neg (\exists x\, A) \fCenter \Delta$
\RightLabel{\nexl}
\UnaryInf$\Gamma, \neg (\exists x\, A) \fCenter \Delta$
\DisplayProof
&
{\em(r24)}
&
\Axiom$\Gamma \fCenter \neg A[x\bs z], \Delta$
\RightLabel{\nexr}
\UnaryInf$\Gamma \fCenter \neg (\exists x\, A), \Delta$
\DisplayProof
&
{\em(r20)}

\end{tabular}
\smallskip

\end{minipage}
}

\caption{\label{fig:seqhat}The sequent calculus LHT for the first-order logic of HT.}
\end{figure}

%
%
%
%



%

\begin{example}
\label{exa:lht}
Consider the formula $F_1 = ( p \impl q ) \mor ( q \impl p )$, which is problem 16 of Pelletier's ``75 problems for theorem proving''~\cite{ar:Pelletier86}. 
%
%
A sequent proof for $F_1$ in LHT is given below, thus $F_1$ is valid in HT. 

\bigskip

\hspace{-1mm}
{
\infer[\orr]{
 \gCenter (p \impl q ) \mor ( q \impl p )
 }
 {\infer[\imr]{
   \gCenter p \impl q ,~ q \impl p
   }
   {\infer[\imr]{ 
     p \gCenter q ,~ q \impl p 
     }
     {\infer[\ax_1]{p, q \gCenter p, q}{}
      &
      \infer[\ax_2]{p, \neg p \gCenter \neg q, q}{}
     }
    &
    \infer[\imr]{ 
     \neg q \gCenter \neg p,~ q \impl p
     }
     {\infer[\ax_2]{\neg q, q \gCenter p, \neg p}{}
      &
      \infer[\ax_1]{\neg q, \neg p \gCenter \neg q, \neg p}{}
     }
   }
 }
}
\bigskip

\noindent
Consider the formula $F_2 = \exists y\, \forall x\,  ( p(y) \impl p(x) )$, which is problem 18 of Pelletier~\cite{ar:Pelletier86}. A sequent proof for $F_2$ in LHT is given below. $ax_1$ and $ax_2$ are $\ax_1$ and $\ax_2$, respectively. For readability, the formula $F_2$ has been omitted from the right side of all four leaf sequents and the formula $p(a)\impl p(b)$ has been omitted from the sequents of the main right branch of the proof.
Thus, $F_2$ is valid in HT.

\bigskip

\hspace*{5.7cm}
{\infer[\!\imr]{}
  {$\small$
   \infer[\!\!\!ax_2\!\!\!\!\!\!]{p(c), \neg p(c)
                      \gCenter \mfn{p(d)}, \mfn{\neg p(b)}}{}
   &
   $\small$
   \infer[\!\!\!ax_1]{\mfn{\neg p(d)}, \neg p(c)
                      \gCenter \neg p(c), \mfn{\neg p(b)}}{}
  }
}
\vspace*{-5mm}

{
\infer[\exr]{
 \gCenter \exists y\, \forall x\, ( p(y) \impl p(x) )
 }
 {\infer[\alr]{
   \gCenter \exists y\,\forall x\, (p(y) \impl p(x))\,,\,
                       \forall x\, (p(a) \impl p(x))
   }
   {\infer[\exr]{
     \gCenter \exists y\,\forall x\, (p(y) \impl p(x))\,,\,
                                      p(a) \impl p(b)
     }
     {\infer[\alr]{
       \gCenter \exists y\,\forall x\, (p(y) \impl p(x))\,,\,
                           \forall x\, (p(b) \impl p(x))\,,\,
                                        p(a) \impl p(b)
       }
       {\infer[\!\!\imr]{
         \gCenter \exists y\,\forall x\, (p(y) \impl p(x))\,,\,
                                          p(b) \impl p(c)\,,\,
                                          p(a) \impl p(b)
         }
         {\infer[\!\!\imr]{
           p(b) \gCenter F_2\,,\,
                         p(a)\impl p(b) \,,\, p(c)
           }
           {$\small$
            \infer[\!\!\!ax_1\!\!\!\!\!\!]{\mfn{p(a)}, p(b) \gCenter
                        p(b),\mfn{p(c)}}{}
            &
            $\small$
            \infer[\!\!\!ax_2\!\!\!\!\!\!]{\neg p(b), p(b) \gCenter
                           \mfn{\neg p(a)}, \mfn{p(c)}}{}
           }
          \hspace*{-1.5mm}
          &
          \infer[\!\!\exr]{
           \neg p(c) \gCenter\exists y\,\forall x\,(p(y)\impl p(x))\,,\,
                             \neg p(b)
           }
           {\infer[\!\!\!\alr]{
             \neg p(c) \gCenter F_2\,,\,
                             \forall x\,(p(c)\impl p(x)) \,,\, \neg p(b)
             }{  \neg p(c) \gCenter F_2\,,\,
                    p(c)\impl p(d) \,,\, \neg p(b)
             }
           }
         }
       }
     }
   }
 }
}
\bigskip

\noindent
%
Neither of the two formulas is valid in intuitionistic logic, but they are both valid in classical logic.


\end{example}

\subsection{Axiomatic Embedding of HT}
\label{ssec:axiomschemas}

The logic of HT can be axiomatized by embedding it into intuitionistic logic. This axiomatization is specified by a set of axioms {$A_1, A_2, \ldots$} that are added to a given formula $F$, such that $F$ is valid in HT iff $(A_1 \mand A_2 \mand \ldots ) \impl F$ is valid in intuitionistic logic.

\begin{definition}[Axiomatic Embedding of HT]
The following two axiom schemas specify the set of axioms of the \emph{axiomatic embedding of HT} into intuitionistic logic~\cite{lipeva07a,ar:Mints10}.

\medskip
\hspace{4.5cm}
$G \mor (G \impl H) \mor \neg H$
\hspace*{3.333cm} (HOS)
\smallskip

\hspace{4.5cm}
$\exists \vx\; ( \,G \impl \forall \vx\; G\, )$
\hspace*{3.9cm} (SQHT)
\medskip

\noindent
$G$ and $H$ are arbitrary (first-order) formulas, which
might contain free variables and $\vx$ is a set $x_1,\ldots,x_n$ of arbitrary variables in $G$.
Furthermore, universal closure is applied to the axiom formulas of these schemas.
More precisely, if $F_A$ is a formula according to one of the axiom schemas and $z_1,\ldots,z_m$ are the free variables in $F_A$, then $\forall z_1, \ldots, z_m\, F_A$ is the resulting axiom.
\end{definition}


Then, a formula $F$ is valid in HT if, and only if, the
formula $F$ with added \emph{HOS} and \emph{SQHT} axioms is valid
in intuitionistic logic.
%
%

\begin{example}
\label{exa:axioms}
For $F_1 = ( p \impl q ) \mor ( q \impl p )$,
the formulas
$( p \mor (p \impl q) \mor \neg q)$ and
$( q \mor (q \impl p) \mor \neg p)$ are instantiations
of the axiom schema HOS.
%
%
%
For $F_2=
\exists y\, \forall x\, ( p(y) \impl p(x) )$ the formulas
$\exists x\, ( p(x) \impl \forall y\, p(y) )$
and
$\exists x\, ( \neg p(x) \impl \forall y\, \neg p(y) )$
are instantiations of the axiom schema SQHT. 
%
\end{example}



\section{Proof Search} 
\label{sec:proofsearch}

The presented sequent calculus LHT and the axiomatic embedding of HT in combination with a proof search calculus for intuitionistic logic allow us to perform \emph{proof search}, i.e.\ finding a proof for the validity of a formula in the logic of HT. For an \emph{efficient} proof search a few additional optimizations are necessary, which are described in this section.

For the axiomatic embedding of HT, we need to determine the validity of a formula in \emph{intuitionistic logic}. One of the most efficient proof methods for intuitionistic logic (see also the results in Section~\ref{sec:evaluation}) is the intuitionistic non-clausal connection calculus, for which we therefore also give a short introduction.

\subsection{Proof Search in the Sequent Calculus LHT}
\label{ssec:optlht}

The following optimizations lead to a more efficient proof search (already) for the \emph{propositional} fragment of the sequent calculus LHT introduced in Section~\ref{ssec:lht}.

\begin{itemize}

\item \textbf{Bottom-up proof search.}
The sequent calculus can be used in a top-down way to derive 
all valid formulas. In order to find a sequent proof for a formula $F$, it is more efficient to start with the sequent $\,\gCenter F$ and apply the rules of LHT in a \emph{bottom-up} (or \emph{analytic}) way until all leaves of the derivation tree end in an axiom of LHT. In the following, we assume such an analytic proof search.

\item \textbf{No backtracking for invertible rules.}
If more than one rule can be applied to a sequent during the proof search, all applicable (alternative) rules have to be applied on backtracking (i.e.\ when a previously applied rule does not lead to a proof). For example, for the sequent $p \mand q \gCenter p \mand q$ the rules $\anl$ as well as $\anr$ are applicable. These choices result in a high amount of backtracking.

This backtracking is not necessary for invertible rules. A rule is called \emph{invertible} iff whenever the conclusion of a rule is provable so are all its premises (the opposite direction obviously holds).

For example, the following (propositional) rules of the intuitionistic sequent calculus LJ~\cite{ar:Gentzen35} are not invertible and, hence, require backtracking (consider, e.g., a proof for the sequent \,$q \gCenter p\mor q$\,):

\medskip
\hspace{1cm}
\Axiom$\Gamma \fCenter A$
\RightLabel{\orr}
\UnaryInf$\Gamma \fCenter A \mor B$
\DisplayProof
\hspace{1.5cm}
\Axiom$\Gamma \fCenter B$
\RightLabel{\orr}
\UnaryInf$\Gamma \fCenter A \mor B$
\DisplayProof
\medskip

All propositional rules of the sequent calculus LHT are invertible (as well as the quantifier rules with Eigenvariable condition marked with *)~\cite{ar:Mints10}. Therefore, no backtracking over alternative rule applications is necessary for these (invertible) rules.

\item \textbf{Apply non-splitting rules first.}
If a sequent rule has more than one premise, the proof search splits into several branches, all of which must be proved. Applying the non-splitting rules first leads to a slightly more efficient proof search as proofs and failed proof attempts are found faster.

\item \textbf{Restricting axiom formulas.}
As in the classical and intuitionistic sequent calculi, the formula $A$ in the axioms can be restricted to literals and atomic formulas. It can be shown that $A$ can be restricted to literals in $\ax_1$ and to atomic formulas in $\ax_2$ (allowing literals for $A$ in $\ax_2$ is redundant as \negg\nel\ and $\ax_1$ could be applied in this case). This leads to two new axioms:

\medskip
\vspace{1mm}
\hspace{1cm}
\AxiomC{}
\RightLabel{\ax$_1$}
\UnaryInfC{$\Gamma, G \fCenter G, \Delta$}
\DisplayProof
\hspace{1.5cm}
\AxiomC{}
\RightLabel{\ax$_2$}
\UnaryInfC{$\Gamma, H, \neg H \fCenter \Delta$}
\DisplayProof
\medskip

where $G$ is a literal and $H$ is an atomic formula. Restricting axiom formulas in this way is necessary when using free variables and unification whenever branches are closed by axioms (see below).

\end{itemize}

These proof search optimizations lead to a \emph{decision procedure} for the \emph{propositional} fragment of HT. This follows from the fact that in each (analytic) proof step the size of at least one formula in a sequent is reduced to one or more smaller formulas, whereas the size and number of all other formulas in the sequent remain unchanged. The \emph{size of a formula} $F$ is its number of atomic formulas and connectives.

\medskip
The first-order logic of HT --- like first-order intuitionistic and first-order classical logic --- is \emph{not} decidable, due to the quantifier rules \all, \exr, \nalr\ and \nexl.
Nevertheless, the following optimizations lead to a more efficient proof search for first-order formulas in the sequent calculus LHT.

\begin{itemize}

\item \textbf{Free variables.}
In order to avoid ``guessing'' an appropriate term $t$ when applying one of the rules \all, \exr, \nalr\ and \nexl\ (which would result in a huge search space), we use free variables instead. In this way we delay the choice for $t$ until axioms are reached, where now term unification is used to make $G$ and $H$ (see above) identical resulting in an appropriate term for $t$.
We replace the rules \all\ and \exr\ by the following new \emph{free variable (quantifier) rules}:

\medskip
\hspace{1cm}
\Axiom$\Gamma, A[x\bs x^*], \forall x\, A \fCenter \Delta$
\RightLabel{\all}
\UnaryInf$\Gamma, \forall x\, A \fCenter \Delta$
\DisplayProof
\hspace{1.5cm}
\Axiom$\Gamma \fCenter \exists x\, A, A[x\bs x^*], \Delta$
\RightLabel{\exr}
\UnaryInf$\Gamma \fCenter \exists x\, A, \Delta$
\DisplayProof
\medskip

where $x^*$ is a new/fresh variable. The rules \nalr\ and \nexl\ are replaced accordingly.

\item \textbf{Skolemization of Eigenvariables.}
In order to check the Eigenvariable condition when using free variables and term unification, a \emph{dynamic skolemization} is used that detects whenever the order of quantifier rule applications violates the Eigenvariable condition (by the occurs check of the term unification). We replace the rules \alr\ and \exl\ by the following new rules:

\medskip
\hspace{1cm}
\Axiom$\Gamma \fCenter A[x\bs t^*], \Delta$
\RightLabel{\alr}
\UnaryInf$\Gamma \fCenter \forall x\, A, \Delta$
\DisplayProof
\hspace{1.5cm}
\Axiom$\Gamma, A[x\bs t^*] \fCenter \Delta$
\RightLabel{\exl}
\UnaryInf$\Gamma, \exists x\, A \fCenter \Delta$
\DisplayProof
\medskip

where $t^*$ is the Skolem term $f^*(x_1,\ldots,x_n)$, in which $f^*$ is a new function symbol and $x_1,\ldots,x_n$ are all free variables in $\Gamma$ and $\Delta$. The rules \nall\ and \nexr\ are replaced accordingly.

%

%

\item \textbf{Iterative deepening.}
The application of the \emph{free variable quantifier} rules (\all, \exr, \nalr\ and \nexl) might lead to non-terminating proof branches. In order to regain completeness, an iterative deepening over the number of free variables that are allowed on a branch in a proof derivation is conducted.
If there are no free variable quantifiers in the given formula, no iterative deepening is conducted leading to a decision procedure for such first-order formulas.

\end{itemize}

\subsection{A Restricted Axiomatic Embedding of HT}
\label{ssec:optaxioms}

In the axiom schemas for HT introduced in Section~\ref{ssec:axiomschemas},
%
%
$G$ and $H$ are arbitrary formulas. This leads to an infinite number of axioms that have to be added (step by step) to the given formula to be proved, resulting in a huge search space.
In order to keep the number of these axioms low, we restrict $G$ to literals and $H$ to atomic formulas. Furthermore, we can restrict the axiom schema \emph{HOS} to formulas $G \neq H$.

\begin{definition}[Restricted Axiomatic Embedding of HT]
Let $F$ be a (first-order) formula, ${\cal P}$ the set of predicate symbols in $F$ and $\langle P\rangle$ the arity of the predicate symbol $P\in {\cal P}$.
The following set of axioms for a formula $F$ specifies a \emph{restricted axiomatic embedding of HT} into intuitionistic logic.

\medskip
\noindent
\hspace*{1mm}
$\mbox{AxiomsHT}:=\{\,\forall x_1,\ldots,x_n\,\forall y_1,\ldots,y_m\, (G \mor (G \impl H) \mor \neg H)\,,~
\exists x_1,\ldots,x_n\, (G \impl \forall x_1,\ldots,x_n\, G ) ~~|~~\\
\hspace*{7mm}
G\,{=}\,P(x_1,...,x_n)$ or $G\,{=}\,\neg P(x_1,...,x_n),\, H\,{=}\,P'(y_1,...,y_m)$ for $P,P' {\in} {\cal P}$ with $P{\neq} P',\, n{=}\langle P\rangle,\, m{=}\langle P'\rangle \}$
%

\end{definition}

The evaluation in Section~\ref{sec:evaluation} reveals whether this is a reasonable restriction.
If both, $G$ and $H$ are restricted to atomic formulas, the weak law of excluded middle $\neg p \mor \neg\neg p$ is not provable any more.
\begin{example}
Consider the formula $F_1 = ( p \impl q ) \mor ( q \impl p )$. The set $\mbox{AxiomsHT}$ includes six axioms (for $G\in \{p,q,\neg p,\neg q\}$ and $H\in\{p,q\}$ with $G\neq H$), of which either $( p \mor (p \impl q) \mor \neg q)$ or $( q \mor (q \impl p) \mor \neg p)$ is necessary and sufficient in order to prove $F_1$ in intuitionistic logic.
%
%
%
%
For the formula $F_2= \exists y\, \forall x\, ( p(y) \impl p(x) )$, the set $\mbox{AxiomsHT}$ includes three axioms (for $G\in \{p(x),\neg p(x)\}$ and $H\in \{p(y)\}$ with $G\neq H$), of which $\exists x\, ( p(x) \impl \forall y\, p(y) )$ is necessary and sufficient to prove $F_2$ in intuitionistic logic.
\end{example}

\subsection{Intuitionistic Non-Clausal Connection Calculus}
\label{ssec:optnanocopi}

%
In contrast to sequent and tableau calculi, which are {\em connective-driven\/}, connection calculi~\cite{bo:Bibel87} use a {\em connection\/}-driven search strategy, leading to
a goal-oriented and, therefore, to a significantly more efficient \emph{proof search}.
A {\em connection\/} is a set $\{{A_1}^0,{A_2}^1\}$ of literals with the same predicate symbol but different polarities (in classical logic polarity 1 would represent a negated formula, representing the set $\{{A_1},\neg {A_2}\}$).
%

The non-clausal connection calculus~\cite{inp:Otten11} works on {\em non-clausal\/} matrices, where a matrix $M$ is a set of clauses and a clause $C$ is a set of literals $L$ \emph{and} (sub)matrices. It represents a formula in negation normal form.
%
A \emph{prefix} is a string consisting of variables ($V$) and constants ($a$) and assigned to each literal. It can be seen as a \emph{world path} that captures the Kripke semantics of intuitionistic logic.

For a formula $F$, polarity $pol\in\{0,1\}$ and prefix $p$, the intuitionistic matrix $M(F^{pol}:p)$ of a prefixed formula $F^{pol}:p$ is defined inductively according to Table~\ref{tab:nonclamat}.
$x^*$ is a new term variable, $t^*$ is the Skolem term $f^*(x_1,\ldots,x_n)$, $V^*$ is a new prefix variable, $a^*$ is the prefix constant of the form $f^*(x_1,\ldots,x_n)$, $f^*$ is a new function symbol and $x_1,\ldots,x_n$ are all free term and prefix variables in the corresponding formula $F^{pol}\,{:}\,p$.
The {\em intuitionistic non-clausal matrix\/} $M(F)$ of a formula $F$ is the intuitionistic matrix $M(F^0:\ep)$.

\begin{table}[hb]
\caption{\label{tab:nonclamat}The (prefixed)
         non-clausal matrix for 
         intuitionistic 
         logic.}
$
\begin{array}[t]{@{}l@{~~~~}l@{~~~~}l@{}}
  \hline
  \mbox{type}\, & \,F^{pol}:p & M(F^{pol}:p)\\
  \hline\hline
  \mbox{\scr atomic} & ~A^0:p & \{\{A^0:p\underline{a^*}\}\} \\
  \hline
  \alpha & (G\wedge H)^1:p & \{\{M(G^1:p)\}\},\{\{M(H^1:p)\}\} \\
         & (G\vee H)^0:p   & \{\{M(G^0:p)\}\},\{\{M(H^0:p)\}\} \\
         & (G\impl H)^0{:}p  & \{\{M(G^1:p\underline{a^*})\}\},\{\{M(H^0:p\underline{a^*})\}\} \\
  \hline
  \beta  & (G\wedge H)^0:p & \{\{M(G^0:p), M(H^0:p)\}\} \\
         & (G\vee H)^1:p   & \{\{M(G^1:p), M(H^1:p)\}\} \\
         & (G\impl H)^1{:}p & \{\{M(G^0:p\underline{V^*}),M(H^1:p\underline{V^*})\}\} \\
  \hline
\end{array}
$
\hspace{5mm}
$
\begin{array}[t]{@{~~}l@{~~~~}l@{~~~~}l@{}}
  \hline
  \mbox{type} & \,F^{pol}:p & M(F^{pol}:p)\\
  \hline\hline
  \mbox{\scr atomic} & ~A^1:p & \{\{A^1:p\underline{V^*}\}\} \\
  \hline
  \alpha & (\neg G)^0:p & M(G^1:p\underline{a^*}) \\
         & (\neg G)^1:p & M(G^0:p\underline{V^*}) \\
  \hline
  \gamma & (\forall x G)^1:p & M(G [x\bs x^*]^1:p\underline{V^*}) \\
         & (\exists x G)^0:p & M(G [x\bs x^*]^0:p)\\
  \hline
  \delta & (\forall x G)^0:p  & M(G [x\bs t^*]^0:p\underline{a^*}) \\
         & (\exists x G )^1:p & M(G [x\bs t^*]^1:p)\\
  \hline
\end{array}
$
\vspace*{3mm}
\end{table}

As usual, a {\em term substitution\/} $\sigma_T$ assigns terms to variables. Furthermore, a {\em prefix substitution\/} $\sigma_P$ assigns strings to
prefix variables and is calculated by a {\em prefix unification}~\cite{inp:Otten17b}.
For intuitionistic logic, a connection $\{A_1^0:p_1,A_2^1:p_2\}$ is {\em $\sigma$-complementary\/} iff $\sigma_T(A_1)=\sigma_T(A_2)$ and $\sigma_P(p_1)=\sigma_P(p_2)$ for a combined substitution $\sigma=(\sigma_T,\sigma_P)$.
The \emph{non-clausal connection calculus} for intuitionistic logic~\cite{inp:Otten17b} is given in Figure~\ref{fig:noclaconcalc}.
An \emph{intuitionistic connection proof} for $F$ is a derivation tree of \,$\varepsilon,M(F),\varepsilon$ in the non-clausal connection calculus, in which all leaves are axioms.
%
%
Compared to the formal {\em clausal\/} connection calculus~\cite{ar:OttenBibel03,inp:OttenBibel17}, the extension rule is generalized
and  a {\em decomposition rule\/} is added~\cite{inp:Otten11,inp:Otten16}.

\begin{figure}[t]
\vspace*{2mm}
\hspace{2.5mm}
\fbox{
\begin{minipage}{15cm}
    \makebox[2.4cm][r]{\em Axiom (A)~~}
    \begin{tabular}[c]{@{}c@{}}
     \\ \hline\\[-3.5mm]
     \hspace*{-0mm}$\{\},M,\mbox{\em Path\/}$\hspace*{-0mm}
    \end{tabular}
\hspace*{18mm}
    \makebox[2.4cm][r]{\em Start (S)~~}
    \begin{tabular}{@{}c@{}}
     \hspace*{-0mm}$C_2,M,\{\}$\hspace*{-0mm}\\ \hline\\[-3.5mm]
     $\varepsilon,\,M,\,\varepsilon$
    \end{tabular}
\hspace*{2mm}
and $C_2$ is copy of $C_1{\in}M$~~
\\[1mm]
    \makebox[2.4cm][r]{\em Reduction (R)~~}
    \begin{tabular}[c]{@{}c@{}}
     $C,M,\mbox{\em Path\/}{\cup}\{L_2\ul{:p_2}\}$\\[0.5mm] \hline\\[-3.5mm]
     \hspace*{-0mm}$C{\cup}\{L_1\,\ul{{:}\,p_1}\},M,
                   \mbox{\em Path}{\cup}\{L_2\,\ul{{:}\,p_2}\}$\hspace*{-0mm}
    \end{tabular}
\hspace*{2mm}
and $\{L_1\ul{{:}p_1},L_2\ul{{:}p_2}\}$ is $\sigma$-complementary
\\[1.5mm]
    \makebox[2.4cm][r]{\em Extension (E)~~}
    \begin{tabular}[c]{@{}c@{}}
     \hspace*{-0mm}$C_3,M[C_1\bs C_2],\mbox{\em Path\/}{\cup}\{L_1\,\ul{{:}\,p_1}\}$\hspace*{0.3cm}
     $C,M,\mbox{\em Path\/}$\hspace*{-0mm}\\[0.5mm] \hline\\[-3.5mm]
     $C{\cup}\{L_1\ul{{:}\,p_1}\},M,\mbox{\em Path\/}$
    \end{tabular}
\hspace*{3mm}
\parbox{5.5cm}{\raggedright
               $C_3{:=}\beta$-$clause_{L_2}(C_2)$, $C_2$ is copy\\ of
               $C_1$,
               $C_1$ is e-clause of $M$ wrt.
               $\mbox{\em Path\/}\,{\cup}\,\{L_1\,\ul{{:}p_1}\}$, $C_2$ contains
               $L_2\,\ul{{:}p_2}$, $\{L_1\ul{{:}p_1},L_2\ul{{:}p_2}\}$ is
               $\sigma$-complementary}
\\[1.5mm]
    \makebox[3.2cm][r]{\em Decomposition (D)~~}
    \begin{tabular}[c]{@{}c@{}}
     $C\cup C_1,M,\mbox{\em Path\/}$\\ \hline\\[-3.5mm]
     $C{\cup}\{M_1\},M,\mbox{\em Path\/}$
    \end{tabular}
\hspace*{2mm}
and $C_1{\in}M_1$
\end{minipage}
}
\caption{\label{fig:noclaconcalc}The non-clausal
         connection calculus for 
           intuitionistic 
           logic.}
\end{figure}

\begin{example}
Consider the two formulas $F_3=(p\impl p)$ and $F_4=(p\mor \neg p)$. Their matrices are $M(F_3)=\{\{p^1:a_1 V_1\},\{p^0:a_1 a_2\}\}$ and $M(F_4)=\{\{p^0:a_1 \},\{p^1:a_2 V_1\}\}$. The only connections in theses matrices are $\{p^1\,{:}\,a_1 V_1, p^0\,{:}\,a_1 a_2\}$ and $\{p^0\,{:}\,a_1 , p^1\,{:}\,a_2 V_1\}$, respectively. The prefixes $a_1 V_1{=}a_1 a_2$ unify under the substitution $\sigma_P(V_1){=}a_2$ whereas $a_1$ and $a_2 V_1$ cannot be unified. Hence, only the first connection is $\sigma$-complementary and, therefore, $F_3$ is intuitionistically valid, whereas $F_4$ is not.
\end{example}

\section{The Implementations}
\label{sec:implementations}

The sequent calculus LHT presented in Sections~\ref{ssec:lht}
with the optimizations described in Section~\ref{ssec:optlht} was implemented in \leanhat. The axiomatic embedding of HT presented in Section~\ref{ssec:axiomschemas} with the optimizations described in Section~\ref{ssec:optaxioms} was implemented in
\ileansephat, \ileantaphat\ and \nanocopihat.%
\footnote{
All presented theorem provers are available 
at \url{https://leancop.de/leanhat/} and \url{https://leancop.de/ihat/}.
}

\subsection{The Sequent Prover \leanHAT}

\leanhat\ is a theorem prover for the first-order logic of HT with equality.
It is a compact Prolog implementation of the sequent calculus LHT presented in Section~\ref{ssec:lht} with the optimizations described in Section~\ref{ssec:optlht}.
The following (Prolog) syntax is used for formulas \texttt{F}: the connectives
``\texttt{,}'' (conjunction), ``\texttt{;}'' (disjunction), ``\verb?~?'' (negation), ``\texttt{=>}'' (implication), ``\texttt{<=>}'' (equivalence), the quantifiers ``\texttt{all X:F}'' (universal) and  ``\texttt{ex X:F}'' (existential); atomic formulas are represented by Prolog terms.
Sequents $\Gamma \gCenter \Delta$ are represented by \texttt{Left > Right} where the multisets \texttt{Left} and \texttt{Right} are represented by Prolog lists of the form \texttt{[F1,\ldots,Fn]}.
The source code of the \leanhat\ theorem prover is shown in Figure~\ref{fig:leanhat}. It is divided into the specification of the sequent rules (lines \emph{r1--r26}) and the actual proof search (lines \emph{1--16}).

\begin{figure}
\vspace*{2mm}
\setlength{\unitlength}{0.01mm}
\hspace*{10mm}
\begin{picture}(13950,0)(0,-70)
\thinlines
{\put(0,0){\line( 0,-1){21100}}
\put(0,-21100){\line( 1, 0){13800}}
\put(13800,-21100){\line( 0, 1){21100}}
\put(13800,0){\line(-1, 0){13800}}
}
\end{picture}

\hspace*{14mm}
\begin{minipage}{7mm}
\vspace*{1mm}\fns
\em
\hspace*{0mm}\\[0mm]
\\[1.3mm]
\hspace*{1.3mm}(r1)\\
\hspace*{1.3mm}(r2)\\
\hspace*{1.3mm}(r3)\\
\hspace*{1.3mm}(r4)\\
\hspace*{1.3mm}(r5)\\
\hspace*{1.3mm}(r6)\\
\hspace*{1.3mm}(r7)\\
\\[-2mm]
\hspace*{1.3mm}(r8)\\
\hspace*{1.3mm}(r9)\\
(r10)\\
(r11)\\
(r12)\\
(r13)\\
(r14)\\
\\[-2mm]
(r15)\\
(r16)\\
(r17)\\
(r18)\\
\\[-2mm]
(r19)\\
(r20)\\
(r21)\\
(r22)\\
\\[-2mm]
(r23)\\
(r24)\\
(r25)\\
(r26)\\
\\[2.3mm]
\hspace*{1.3mm}~~(a)\\
\hspace*{1.3mm}~~(b)\\
\\[0mm]
\\[0mm]
\\[0mm]
\hspace*{1.3mm}~~(1)\\
\hspace*{1.3mm}~~(2)\\
\hspace*{1.3mm}~~(3)\\
\hspace*{1.3mm}~~(4)\\
\\[0mm]
\hspace*{1.3mm}~~(5)\\
\hspace*{1.3mm}~~(6)\\
\hspace*{1.3mm}~~(7)\\
\hspace*{1.3mm}~~(8)\\
\hspace*{1.3mm}~~(9)\\
\hspace*{0mm}~~(10)\\
\hspace*{0mm}~~(11)\\
\hspace*{0mm}~~(12)\\
\hspace*{0mm}~~(13)\\
\hspace*{0mm}~~(14)\\
\hspace*{0mm}~~(15)\\
\hspace*{0mm}~~(16)
\end{minipage}
\hspace{6mm}
\begin{minipage}{12.2cm}\fns
\vspace*{1.5mm}
\begin{alltt}
% - specification of sequent rules

fml((A,B),    1, [A,B] > [],   [] > [],     [] > [], _,_,[],[],[]).
fml((A;B),    0, [] > [A,B],   [] > [],     [] > [], _,_,[],[],[]).
fml(~ (A,B),  0, [] > [~A,~B], [] > [],     [] > [], _,_,[],[],[]).
fml(~ (A;B),  1, [~A,~B] > [], [] > [],     [] > [], _,_,[],[],[]).
fml(~ (A=>B), 1, [~B] > [~A],  [] > [],     [] > [], _,_,[],[],[]).
fml(~ (~A),   1, [] > [~A],    [] > [],     [] > [], _,_,[],[],[]).
fml(~ (~A),   0, [~A] > [],    [] > [],     [] > [], _,_,[],[],[]).
\end{alltt}
\vspace*{-3.8mm}
\begin{alltt}
fml((A,B),    0, [] > [A],     [] > [B],    [] > [], _,_,[],[],[]).
fml((A;B),    1, [A] > [],     [B] > [],    [] > [], _,_,[],[],[]).
fml(~ (A,B),  1, [~A] > [],    [~B] > [],   [] > [], _,_,[],[],[]).
fml(~ (A;B),  0, [] > [~A],    [] > [~B],   [] > [], _,_,[],[],[]).
fml(~ (A=>B), 0, [~A] > [],    [] > [~B],   [] > [], _,_,[],[],[]).
fml((A=>B),   0, [A] > [B],    [~B] > [~A], [] > [], _,_,[],[],[]).
fml((A=>B),   1, [~A] > [],    [] > [A,~B], [B]>[],  _,_,[],[],[]).
\end{alltt}
\vspace*{-3.8mm}
\begin{alltt}
fml((A<=>B),  1, [((A=>B),(B=>A))] > [],[]>[],[]>[], _,_,[],[],[]).
fml((A<=>B),  0, [] > [((A=>B),(B=>A))],[]>[],[]>[], _,_,[],[],[]).
fml(~ (A<=>B),1, [~ ((A=>B),(B=>A))]>[],[]>[],[]>[], _,_,[],[],[]).
fml(~ (A<=>B),0, []>[~ ((A=>B),(B=>A))],[]>[],[]>[], _,_,[],[],[]).
\end{alltt}
\vspace*{-3.8mm}
\begin{alltt}
fml(~all X:A, 1, [~C] > [], []>[], []>[],  S,FV,[],(X,A),(S^FV,C)).
fml(~ex X:A,  0, [] > [~C], []>[], []>[],  S,FV,[],(X,A),(S^FV,C)).
fml( all X:A, 0, [] > [C],  []>[], []>[],  S,FV,[],(X,A),(S^FV,C)).
fml( ex X:A,  1, [C] > [],  []>[], []>[],  S,FV,[],(X,A),(S^FV,C)).
\end{alltt}
\vspace*{-3.8mm}
\begin{alltt}
fml(~all X:A, 0, [] > [~C,~all X:A], []>[],[]>[], _,_,[Y],X:A,Y:C).
fml(~ex X:A,  1, [~C,~ex X:A] > [],  []>[],[]>[], _,_,[Y],X:A,Y:C).
fml( all X:A  1, [C,all X:A] > [],   []>[],[]>[], _,_,[Y],X:A,Y:C).
fml( ex X:A,  0, [] > [C,ex X:A],    []>[],[]>[], _,_,[Y],X:A,Y:C).
\end{alltt}
\vspace*{0.5mm}
\begin{alltt}
prove(F,I) :- prove([]>[F],s,[],I).
prove(F,I) :- \verb?\?+ no_mul([([F],0)]), I1 is I+1, prove(F,I1).

% - proof search

prove(Left>Right,_,_,_) :-
    member(A,Left), ( member(B,Right) ; member(~B,Left) ), ( A==B ->
    ! ; ( A=(~C) ->  C=..[P|_] ; A=..[P|_] ), P\verb?\?=',', P\verb?\?=';', P\verb?\?='~',
    P\verb?\?='=>', P\verb?\?='<=>', P\verb?\?=all, P\verb?\?=ex, unify_with_occurs_check(A,B) ).

prove(Left>Right,S,FreeV,VarLim) :-
    fml(F,Pol,L1>R1,L2>R2,L3>R3,S,FreeV,V,Cpy1,Cpy2),
    ( Pol=1, append(LeftA,[F|LeftB],Left), RightA=Right, RightB=[] ;
      Pol=0, append(RightA,[F|RightB],Right), LeftA=Left, LeftB=[] ),
    ( V=[] -> ! ; \verb?\?+ length(FreeV,VarLim) ),
    append(FreeV,V,FreeV1), copy_term((Cpy1,FreeV),(Cpy2,FreeV)),
    append(LeftA,LeftB,L0), append(RightA,RightB,R0),
    append(L1,L0,Left1), append(R1,R0,Right1), append(L2,L0,Left2),
    append(R2,R0,Right2), append(L3,L0,Left3), append(R3,R0,Right3),
    prove(Left1>Right1,l(S),FreeV1,VarLim),
    ( L2=[],R2=[] -> true ; prove(Left2>Right2,r(S),FreeV1,VarLim) ),
    ( L3=[],R3=[] -> true ; prove(Left3>Right3,x(S),FreeV1,VarLim) ).
\end{alltt}
\vspace{-1.7mm}
\end{minipage}

\caption{\label{fig:leanhat}The source code of the \leanhat\ sequent prover for the logic of HT.}
\end{figure}

%
%
%

The predicate
\texttt{fml(F,Pol,L1>R1,L2>R2,L3>R3,S,FreeV,NewX,Cpy1,Cpy2)}
is used to specify the rules of the sequent calculus LHT in Figure~\ref{fig:seqhat}.
Each of the 22 rules of the calculus corresponds to exactly one Prolog clause of the predicate \texttt{fml}, whose number is given in parentheses after the name of the rule. Furthermore, there are four additional clauses for equivalences $A \equi B$ (lines \emph{r15--r18}) defined as $(A \impl B) \mand (B \impl A)$. Non-splitting rules are applied first, as are quantifier rules with Eigenvariables.
\texttt{F} is the main formula of the rule's conclusion, \texttt{Pol} is its polarity, which is either 0 or 1 (i.e.\ right or left side of the sequent), \texttt{L$n$>R$n$} (for $n=1,2,3$) are the formulas that are added to the $n$-th premise of the rule, \texttt{S} is a unique position used as Skolem functor, \texttt{FreeV} is a list of free variables in the sequent, \texttt{NewX} is the new free variable (or the empty list, otherwise), \texttt{Cpy1} is copied and unified with \texttt{Cpy2}.

The actual proof search is done by the Prolog predicate 
\texttt{prove(Left>Right,S,FreeV,VarLim)}. It succeeds iff there is a sequent proof for the sequent \texttt{Left>Right} with the current position \texttt{S} and the current list of free variables \texttt{FreeV} using not more than \texttt{VarLim} free variables in every proof branch.

The two axioms of the sequent calculus LHT are implemented in the first four lines (lines \emph{1--4}). Sound unification of the two axiom literals is performed (as free variables are used). This is done by the built-in predicate \texttt{unify_with_occurs_check} (line \emph{4}).
If none of the axioms can be applied, a rule of the calculus is picked (line \emph{6}) and checked if it can be applied, either to the left side (line \emph{7}) or the right side (line \emph{8}) of the current sequent. If an invertible rule is applied, any backtracking is cut off (using the Prolog cut ``\texttt{!}''), otherwise it is checked whether the current number of free variables is within the current limit \texttt{VarLim} (line \emph{9}).

Afterwards, the new free variable (if any) is added to the list of free variables and a copy of \texttt{Cpy1} is made and unified with \texttt{Cpy2} using the built-in predicate \texttt{copy\_term} (line \emph{10}). These copies are necessary whenever a quantifier rule is applied: (1) For the free variable quantifier rules \all, \exr, \nalr\ and \nexl, a fresh copy of the main formula has to be added to the sequent of the premise; and (2) for the Eigenvariable quantifier rules \alr, \exl, \nall and \nexr, a Skolem term has to be inserted for the Eigenvariable. In order to avoid that free variables are renamed, the list of free variables \texttt{FreeV} is added as an ``argument'' on both sides of the \texttt{copy\_term} predicate.
The next three lines add any new formulas \texttt{R}$i$ and \texttt{L}$i$ (for $i=1,2,3$) to the right and left side of the $i$-th premise, respectively (lines \emph{11--13}). Afterwards the proof search continues with the first premise (line \emph{14}), (if existent) the second premise (line \emph{15}) and (if existent) the third premise (line \emph{16}).



The main predicate \texttt{prove(F,I)} realizes iterative deepening and succeeds iff there is a sequent proof for the formula \texttt{F}. The integer \texttt{I} is the (initial) limit for the number of free variables on each branch in the sequent calculus. This predicate invokes the actual proof search by calling \texttt{prove([]>[F],s,[],I)} with the start sequent \texttt{[]>[F]}, the start position \texttt{s}, the empty list \texttt{[]} of free variables and the free variable limit \texttt{I} (line \emph{a}). If this call fails, the free variable limit is increased and the proof search starts again (line \emph{b}). The predicate \texttt{no_mul([([F],0)])} succeeds iff there are no free variable quantifiers in \texttt{F}, i.e.\ quantifiers that lead to the application of a free variable quantifier rule, in which case the iterative deepening does \emph{not} continue (line \emph{b}) (``\verb?\?\texttt{+}'' is the Prolog negation). The \texttt{no_mul} predicate is implemented in six lines of code and the only code of the \leanhat\ (core) prover not included in Figure~\ref{fig:leanhat}.
Therefore, \leanhat\ is (also) a decision procedure for first-order formulas that do not contain free variable quantifiers and therefore also for the propositional fragment of HT.
%
%
%

If the given formula includes equality, the necessary axioms are added in a preprocessing step.

\begin{example}[Sequent prover leanHaT]
Consider again the three formulas
$F_1 = ( p \impl q ) \mor ( q \impl p )$,
$F_2 = \exists y\, \forall x\, ( p(y) \impl p(x) )$
\,and\, $F_4=(p\mor \neg p)$.
In the ILTP problem library~\cite{ar:RathsOttenKreitz07} (see also Section~\ref{ssec:evaliltp}), these formulas are included as problems SYN416+1, SYN048+1 and SYN387+1, respectively.
In order to determine whether these formulas are valid, we load the \leanhat\ prover into Prolog (using \textup{\texttt{[leanhat].}}) and then invoke the following goals: 
\textup{\texttt{prove( ( (p=>q) ; (q=>p) ), 1)}}\,,\,
\textup{\texttt{prove( ( ex Y: all X: ( p(Y) => p(X) ) ), 1)}}
\,and\,
\textup{\texttt{prove( ( p ; }}\verb?~?\textup{\texttt{ p ), 1)}}\,.
For the first two formulas Prolog returns
``\textup{\texttt{Yes}}'', for the third formula it returns
``\textup{\texttt{No}}''. Therefore, formulas $F_1$ and $F_2$ are valid in the logic of HT, whereas $F_4$ is not.
\end{example}

\subsection{The Axiomatic Sequent and Tableau Provers \ileansephat\ and \ileanTAPhat}

The theorem prover \ileansephat\ is an axiomatic extension of the intuitionistic sequent prover \ileansep\ for the first-order logic of HT with equality.
\ileansep\ is a compact Prolog implementation of the single-succedent intuitionistic sequent calculus.
Similar to tableau calculi, it uses a bottom-up proof search, free variables and a dynamic skolemization to deal with quantifiers.
Together with the occurs check of term unification, this ensures that the Eigenvariable condition is respected.
If necessary, equality axioms are added in a preprocessing step and  treated with regard to the logic of HT.


The theorem prover \,\ileantaphat\, is an axiomatic extension of the intuitionistic tableau prover \ileantap\ for the first-order logic of HT with equality.
\ileantap~\cite{inp:Otten97} is a compact Prolog implementation of a prefixed tableau calculus, similar to Fitting's prefixed tableau calculi~\cite{bo:Fitting83}. It extends the classical calculus~\cite{bo:Smullyan68} by adding prefixes to capture the Kripke semantics of intuitionistic logic, uses free prefix variables~\cite{bo:Wallen90} and extends skolemization to prefix constants.
First, \ileantaphat\ performs a classical proof search collecting prefixes of literals that close branches. If this search succeeds, a prefix unification is used to unify these prefixes and ensure that the given formula is also intuitionistically valid.
%
%
%
If necessary, equality axioms are added in a preprocessing step and  treated with regard to HT.

Both provers for HT are likely incomplete, i.e.\ there are valid formulas for which they cannot find proofs, as the axiomatization is optimized with respect to the number of added axioms (see Section~\ref{ssec:optaxioms}).

\subsection{The Axiomatic Connection Prover \nanocopihat}

The automated theorem prover \nanocopihat\ is an axiomatic extension of the intuitionistic non-clausal connection prover \nanocopi\ for the first-order logic of HT with equality.
\nanocopi\ is a 
compact Prolog implementation of the non-clausal connection calculus for {\em intuitionistic\/} first-order  logic~\cite{inp:Otten17b,inp:Otten21} and an extension of the classical prover \nanocop~\cite{inp:Otten16}. 
It is based on the prefixed non-clausal connection calculus presented in Section~\ref{ssec:optnanocopi}.
First, \nanocopi\ performs a classical proof search, in which the prefixes of each connection are collected.
If this search succeeds, the prefixes of the literals in each connection are unified.
Additional optimization techniques are regularity, lemmata, restricted backtracking and strategy scheduling~\cite{ar:Otten10,inp:Otten21}.
If necessary, equality axioms are added in a preprocessing step and  treated with regard to the logic of HT.
Currently, \nanocopi~2.0 is one of the fastest theorem provers for first-order intuitionistic logic~\cite{inp:Otten21}.
Again, \nanocopihat\ is likely incomplete as the embedding into intuitionistic logic uses only a restricted number of axioms as specified in Section~\ref{ssec:optaxioms}.%
\footnote{
The focus of the presented proof calculi, methods and provers is to \emph{prove} formulas. Effectively \emph{refuting} formulas requires
different or additional methods and techniques.
For this reason, completeness (of methods and provers) is less relevant.
For example, \nanocopihat\ proves significant more problems than \leanhat\ (see Section~\ref{ssec:evalresults}) even though it is (likely) incomplete. Furthermore, one of the most successful (proof search) techniques used in \nanocopi\ (and therefore also by \nanocopihat) is incomplete~\cite{inp:Otten21,ar:Otten10} (overall, \nanocopi\ is complete as complete strategies are also part of the used strategy scheduling).
Finally, just as first-order classical and first-order intuitionistic logic, the first-order logic of HT is also not decidable.
%
}

\section{Experimental Evaluation} 
\label{sec:evaluation}

All four implementations presented in Section \ref{sec:implementations}
were evaluated on a large set of problems of an existing benchmark library and also compared to theorem provers for intuitionistic and classical logic.

\subsection{Benchmark Problems}
\label{ssec:evaliltp}

%

\begin{figure}[t]
\vspace{-1.5cm}
\hspace{6mm}
\begin{picture}(200,54)(230,50)
\thicklines
\setlength{\unitlength}{1.25mm}
\put (88,10){\oval(56,12)}
\put (81,10){\oval(37,10)}
\put (75,10){\oval(20, 8)}
\put (67,11){intuitionistic}
\put (69, 7){theorems}
\put (90,11){HT}
\put (87, 7){theorems}
\put (103,11){classical}
\put (103, 7){theorems}
\end{picture}
\hspace{5mm}
\begin{picture}(200,54)(230,50)
\thicklines
\setlength{\unitlength}{1.25mm}
\put (92,10){\oval(64,12)}
\put (83,10){\oval(41,10)}
\put (75,10){\oval(20, 8)}
\put (69,11){classical}
\put (66.5, 7){non-theorems}
\put (92,11){HT}
\put (85.5, 7){non-theorems}
\put (105,11){intuitionistic}
\put (105, 7){non-theorems}
\end{picture}
\vspace{1.1cm}
\caption{Relation between intuitionistic logic, HT and classical logic.}
\label{fig:evalrelation}
\end{figure}
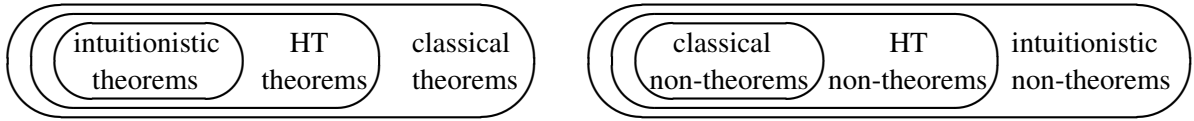

At present, larger sets of formulas for testing and benchmarking theorem provers for the first-order logic of HT are not available. So far, there exists only a small set of problems representing strong equivalences of logic programs developed in the context of the \emph{anthem} system~\cite{falilusc20a,faglhaheliliscst25a}. As the syntax of HT is the same as for classical and intuitionistic logic, we can use existing benchmark formulas for these logics, even though the results on these formulas do not have any obvious meaning.
As the logic of HT is an extension of intuitionistic logic, we use the ILTP problem library~\cite{ar:RathsOttenKreitz07} for intuitionistic logic to test and benchmark our four theorem provers for HT. The relation between HT, intuitionistic logic and classical logic is shown in Figure~\ref{fig:evalrelation}.

Version 1.1.2 of the ILTP problem library contains 2550 first-order formulas with status and difficulty rating information for both classical and intuitionistic logic.
The problems are divided into 24 categories and taken from a wide range of different domains,
such as agents~(AGT), general algebra~(ALG), constructive geometry~(GEJ), geometry~(GEO), knowledge representation~(KRS), management~(MGT), natural language processing~(NLP), number theory~(NUM), set theory~(SET), software verification~(SWV), syntactic~(SYN) and intuitionistic syntactic~(SYJ).

The problems in the ILTP library are in TPTP syntax and consist of one \mbox{\em Conjecture} formula and (optionally) $n$ Axiom formulas $\mbox{\em Axiom}_1, \ldots, \mbox{\em Axiom}_n$ ($n \geq 0$). Then, the resulting formula to be proven is
$(\mbox{\em Axiom}_1 \mand \ldots \mand \mbox{\em Axiom}_n) \impl \mbox{\em Conjecture}$. If there is no \mbox{\em Conjecture} (formula), then 
the unsatisfiability of the formula $(\mbox{\em Axiom}_1 \mand \ldots \mand \mbox{\em Axiom}_n)$ has to be determined.

\subsection{Evaluation}
\label{ssec:evalresults}


All four theorem provers for HT described in Section~\ref{sec:implementations} were evaluated on the 2550 problems of the ILTP library.
The tests were conducted on a LIFEBOOK U9311 with a 4-core i7-1185G7 CPU and 16GB of RAM, running Linux 5.15.0 and ECLiPSe Prolog version 5.10.%
\footnote{
ECLiPSe Prolog 5.x is available at
\url{https://eclipseclp.org/Distribution/Builds/};
newer versions of ECLiPSe Prolog are missing important features and have a significantly lower performance.
}

Table~\ref{tab:perform-iltp-ht} shows the results of the evaluation on all 2550 first-order problems of the ILTP library~v1.1.2 for a CPU time limit of 10 seconds.%
\footnote{
Due to the time complexity of the underlying proof search problem, a higher time limit would only have a small impact on the number of proved problems/formulas. Only the \nanocopi\ and \nanocopihat\ provers would benefit a little from a higher time limit, as the strategy scheduling would then be able to execute all 12 (built-in) strategies instead of just the first nine.
}
Included are the HT provers \leanhat~1.0, \ileansephat~1.0, \ileantaphat~1.0 and \nanocopihat~2.0. Also included are the intuitionistic provers \ileansep~1.0, \ileantap~1.17~\cite{inp:Otten97} and \nanocopi~2.0~\cite{inp:Otten21}. The intuitionistic provers provide a lower limit for a (native) theorem prover for HT using a similar calculus. For example, a (native) sequent or tableau prover for HT should solve at least as many problems as \ileansep, a (native) connection prover for HT at least as many problems as \nanocopi.
The table shows for each prover the total number of proved problems,
the number of problems proved within a certain time, the number of proved problems for the most important domains (see Section~\ref{ssec:evaliltp}),
the number of ``unique'' problems proved only by this prover (with respect to its logic), the number of refuted (invalid) problems and the number of errors (usually an overflow of the global/trail stack
of ECLiPSe Prolog). As the axiomatic provers for HT are likely incomplete, they do not refute any (invalid) problems.

Of the HT provers, \nanocopihat\ proves by far the highest number of problems. The sequent prover \leanhat\ comes in second, whereas the \ileansep\ and \ileantap-based provers perform significantly worse.
Both \nanocopihat\ and \leanhat\ prove the highest number of unique problems, subsuming \ileansephat.
%
%
By adding the conjecture start clause technique to the third strategy of \nanocopihat\ (i.e.\ using \texttt{[conj,scut,cut]} instead of \texttt{[scut,cut]}, see~\cite{inp:Otten21}), the number of proved problems increases to 635.
Altogether, \nanocopihat\ and \leanhat\ prove a total of 652 problems.



%

\begin{table}[t]
\caption{Results of the HT theorem provers on the ILTP library.}
\label{tab:perform-iltp-ht}

\begin{center}

\begin{tabular}[c]{@{}l@{}r@{~~~}r@{~~~}r@{~~~}r@{~~~}r@{~~~}r@{~~~}r@{}}
\hline

{\bf Prover} & \ileantap & \ileantaphat & \ileansep & \ileansephat 
             & \leanhat  & \nanocopi    & \nanocopihat\\
   Logic & IL & \HT & IL & \HT & \HT & IL & \HT\\
\hline
{\bf proved}
         & 312 & 154 & 300 & 209 & {\bf 378}& 799 & {\bf 626}\\
\hline
~~0 to ~~~1sec.
         & 306 & 145 & 273 & 184 & {\bf 345}& 699 & {\bf 528}\\
~~1 to ~10sec.
         &   6 &   9 &  27 &  25 & {\bf  33}& 100 & {\bf  98}\\
\hline
AGT &  0 &  0 &  4 & 3 &  4 & 15 & 12 \\
ALG &  6 &  6 &  0 & 0 &  3 & 12 &  9 \\
GEJ & 11 &  0 &  7 & 3 & 11 & 85 & 59 \\
GEO &  0 &  0 &  1 & 0 &  1 & 10 &  7 \\
KRS & 18 &  0 & 17 & 3 & 31 & 91 & 41 \\
MGT &  6 &  0 &  1 & 0 &  1 & 32 & 16 \\
NLP & 11 & 11 &  3 & 0 &  1 & 13 &  8 \\
NUM &  1 &  0 &  0 & 0 &  0 & 49 & 30 \\
SET & 14 &  1 & 19 &12 & 16 &112 & 76 \\
SWV & 51 & 48 & 82 &71 &107 &134 &127 \\
SYN &107 & 56 &102 &93 &134 &125 &139 \\
SYJ & 81 & 28 & 62 &23 & 67 &102 & 90 \\
\hline
unique & 0 & 3 & 2 & 0 & {\bf 24} & 413 & {\bf 238}  \\
\hline
{\bf refuted}
         &   4 &   0 &   4 &   0 & {\bf  31}&  88 & {\bf 0}\\
error    &  60 &  27 & 151 & 114 & {\bf   0}&   3 & {\bf 3}\\
\hline
\end{tabular}

\end{center}
\end{table}

Theorem provers for HT can also be used to refute problems in intuitionistic logic (if $F$ is not valid in HT, then $F$ is also not valid in intuitionistic logic). For example, some of the 18 scalable (and \emph{classically} valid) problems SYJ207 to SYJ212 in the ILTP library are hard to refute even for \emph{propositional} intuitionistic provers. \leanhat\ easily refutes all of these problems within a fraction of a second.

Table~\ref{tab:perform-iltp-c} shows the results of five theorem provers for classical logic: the (native) classical provers \leantap~2.3~\cite{ar:BeckertPosegga95} and \nanocop~2.0~\cite{ar:Otten10}, and the three intuitionistic provers \ileantap, \ileansep\ and \nanocopi\ using an embedding of classical into intuitionistic logic by adding the axiom schema $A \mor \neg A$ for the law of excluded middle (for all atomic formulas $A$).
They provide an upper limit for axiomatic approaches. For example, the performance of \nanocopi-C is the best we can expect when using the axiomatic embedding of HT in combination with the \nanocopi\ prover.



\begin{table}[b]
\caption{Results of the classical theorem provers on the ILTP library.}\label{tab:perform-iltp-c}

\begin{center}

\begin{tabular}[c]{@{}lrrrrr@{}}
\hline

{\bf Prover} & ~~\ileantap-C & ~~\ileansep-C & ~~~~~\leantap
             & \nanocopi-C & ~~~\nanocop\\
   Logic & CL & CL & CL & CL & CL\\
\hline
{\bf proved}
         & 252 & 274 & 516 & 855 & 1086\\
\hline
~~0 to ~~~1sec.
         & 242 & 248 & 499 & 728 &  942\\
~~1 to ~10sec.
         &  10 &  26 &  17 & 127 &  144\\
\hline
{\bf refuted}
         &   0 &   0 &   0 &   5 &   86\\
error    &  82 & 105 & 375 &   3 &    1\\
\hline
\end{tabular}

\end{center}
\end{table}

\section{Conclusion} 



Even though the logic of HT provides the logical foundation of ASP, only little effort has been done in developing tools and efficient calculi that automate formal reasoning in HT. This paper addresses this issue by introducing four automated theorem provers for the \emph{first-order} logic of HT: \leanhat\ implements an optimized native sequent calculus for HT;
\ileansephat, \ileantaphat\ and \nanocopihat\ implement axiomatic embeddings into intuitionistic logic in combination with the intuitionistic theorem provers \ileansep, \ileantap\ and \nanocopi.
To the best of our knowledge, these are
the first theorem provers for the \emph{first-order} logic of HT that do not require a translation of the given formula.

It is notable that all four presented theorem provers are based on \emph{non-clausal} proof calculi, i.e. they do not require the translation of the given formula into a (conjunctive or disjunctive) clausal form, but preserve the structure of the original formula throughout the proof search process. This is in particular important if readable proofs are to be returned.

All four provers for HT were evaluated on the ILTP problem library providing comprehensive performance results. \leanhat\ proves more problems than similar sequent or tableau provers for intuitionistic logic. \leanhat\ solves (i.e.\ proves and refutes) about 20\% fewer problems than the \emph{classical} tableau prover \leantap; a solid result, considering that some formulas proved by \leantap\ are not valid in HT.

For the axiomatic embedding, a large number of axioms need to be added to the given formula. For example, for a propositional formula with 25 distinct predicate symbols, more than 1200 axioms are added, which already takes our optimized axiomatization into account.
As expected and seen with similar approaches, these additional axioms increase the search space significantly. Only in combination with the intuitionistic connection prover \nanocopi\ this approach leads to a convincing performance as shown by the results of the \nanocopihat\ prover.

It is still an open question, which axioms are necessary for a complete and sound axiomatic embedding into intuitionistic logic. We did not find any evidence that our optimized axiomatization is incomplete for \emph{propositional} HT. For \emph{first-order} HT it is likely incomplete. For example, the formula $\exists y\, ( \exists x\, p(x) \impl p(y) )$ (problem SYN971+1 in the ILTP library) can easily be proved by leanHaT, 
but not by any of the axiomatic theorem provers for HT. 
Further theoretical and practical investigations are necessary to answer this open question.%
%




Although the performance results of \leanhat\ and \nanocopihat\ are very promising, the main focus of the research presented in this paper is not efficiency, but to initiate the development of theorem provers for the first-order logic of HT and providing a foundation for future developments of more efficient theorem provers.
Neither (native) sequent or tableau calculi nor axiomatic embeddings are a basis for efficient theorem provers for HT. To this end, native proof methods that are based on a goal-oriented proof search, such as connection calculi need to be developed. This will be part of future work.

A comparison with the \emph{anthem} system and its translation of HT into first-order classical logic will be part of future work as well.
Furthermore, the development of a benchmark library for HT that also contains  problems from real applications, e.g., representing strong equivalences of logic programs would be very useful and would further support the development of efficient theorem provers for HT.
%








%

\section*{Acknowledgments}

The authors would like to thank the members of the \emph{anthem} group at the University of Potsdam, the University of Nebraska at Omaha, and the University of Texas at Austin for many inspiring discussions.
Thanks also goes to the anonymous reviewers for their helpful comments.


\bibliographystyle{eptcs}

\bibliography{iclp25}

\begin{thebibliography}{10}
\providecommand{\bibitemdeclare}[2]{}
\providecommand{\surnamestart}{}
\providecommand{\surnameend}{}
\providecommand{\urlprefix}{Available at }
\providecommand{\url}[1]{\texttt{#1}}
\providecommand{\href}[2]{\texttt{#2}}
\providecommand{\urlalt}[2]{\href{#1}{#2}}
\providecommand{\doi}[1]{doi:\urlalt{https://doi.org/#1}{#1}}
\providecommand{\eprint}[1]{arXiv:\urlalt{https://arxiv.org/abs/#1}{#1}}
\providecommand{\bibinfo}[2]{#2}

\bibitemdeclare{article}{agcadipevi13a}
\bibitem{agcadipevi13a}
\bibinfo{author}{Felicidad \surnamestart Aguado\surnameend},
  \bibinfo{author}{Pedro \surnamestart Cabalar\surnameend},
  \bibinfo{author}{Martin \surnamestart Di{\'e}guez\surnameend},
  \bibinfo{author}{Gilberto \surnamestart P{\'e}rez\surnameend} \&
  \bibinfo{author}{Concepci{\'o}n \surnamestart Vidal\surnameend}
  (\bibinfo{year}{2013}): \emph{\bibinfo{title}{Temporal\, equilibrium\,
  logic:\, a survey}}.
\newblock {\slshape \bibinfo{journal}{Journal of Applied Non-Classical Logics}}
  \bibinfo{volume}{23}(\bibinfo{number}{1--2}), pp. \bibinfo{pages}{2--24},
  \doi{10.1080/11663081.2013.798985}.

\bibitemdeclare{inproceedings}{bedaei16a}
\bibitem{bedaei16a}
\bibinfo{author}{Harald \surnamestart Beck\surnameend}, \bibinfo{author}{Minh
  \surnamestart Dao-Tran\surnameend} \& \bibinfo{author}{Thomas \surnamestart
  Eiter\surnameend} (\bibinfo{year}{2016}): \emph{\bibinfo{title}{Equivalent
  Stream Reasoning Programs}}.
\newblock In \bibinfo{editor}{Kambhampati}  \cite{ijcai16}, pp.
  \bibinfo{pages}{929--935}.

\bibitemdeclare{article}{ar:BeckertPosegga95}
\bibitem{ar:BeckertPosegga95}
\bibinfo{author}{Bernhard \surnamestart Beckert\surnameend} \&
  \bibinfo{author}{Joachim \surnamestart Posegga\surnameend}
  (\bibinfo{year}{1995}): \emph{\bibinfo{title}{lean{TAP}: Lean Tableau-based
  Deduction}}.
\newblock {\slshape \bibinfo{journal}{Journal of Automated Reasoning}}
  \bibinfo{volume}{15}(\bibinfo{number}{3}), pp. \bibinfo{pages}{339--358},
  \doi{10.1007/BF00881804}.

\bibitemdeclare{book}{bo:Bibel87}
\bibitem{bo:Bibel87}
\bibinfo{author}{Wolfgang \surnamestart Bibel\surnameend}
  (\bibinfo{year}{1987}): \emph{\bibinfo{title}{Automated Theorem Proving}}.
\newblock \bibinfo{series}{Artificial intelligence}, \bibinfo{publisher}{F.
  Vieweg und Sohn}, \bibinfo{address}{Wiesbaden},
  \doi{10.1007/978-3-322-90102-6}.

\bibitemdeclare{article}{cafafa20a}
\bibitem{cafafa20a}
\bibinfo{author}{Pedro \surnamestart Cabalar\surnameend},
  \bibinfo{author}{Jorge \surnamestart Fandinno\surnameend} \&
  \bibinfo{author}{Luis \surnamestart {Fari{\~n}as del Cerro}\surnameend}
  (\bibinfo{year}{2020}): \emph{\bibinfo{title}{Autoepistemic Answer Set
  Programming}}.
\newblock {\slshape \bibinfo{journal}{Artificial Intelligence}}
  \bibinfo{volume}{289}, p. \bibinfo{pages}{103382},
  \doi{10.1016/j.artint.2020.103382}.

\bibitemdeclare{inproceedings}{cakaossc16a}
\bibitem{cakaossc16a}
\bibinfo{author}{Pedro \surnamestart Cabalar\surnameend},
  \bibinfo{author}{Roland \surnamestart Kaminski\surnameend},
  \bibinfo{author}{Max \surnamestart Ostrowski\surnameend} \&
  \bibinfo{author}{Torsten \surnamestart Schaub\surnameend}
  (\bibinfo{year}{2016}): \emph{\bibinfo{title}{An {ASP} Semantics for Default
  Reasoning with Constraints}}.
\newblock In \bibinfo{editor}{Kambhampati}  \cite{ijcai16}, pp.
  \bibinfo{pages}{1015--1021}.

\bibitemdeclare{misc}{faglhaheliliscst25a}
\bibitem{faglhaheliliscst25a}
\bibinfo{author}{Jorge \surnamestart Fandinno\surnameend},
  \bibinfo{author}{Christoph \surnamestart Glinzer\surnameend},
  \bibinfo{author}{Zachary \surnamestart Hansen\surnameend},
  \bibinfo{author}{Jan \surnamestart Heuer\surnameend}, \bibinfo{author}{Yuliya
  \surnamestart Lierler\surnameend}, \bibinfo{author}{Vladimir \surnamestart
  Lifschitz\surnameend}, \bibinfo{author}{Torsten \surnamestart
  Schaub\surnameend} \& \bibinfo{author}{Tobias \surnamestart
  Stolzmann\surnameend} (\bibinfo{year}{2025}): \emph{\bibinfo{title}{Anthem
  2.0: Automated Reasoning for Answer Set Programming}}.
\newblock \urlprefix\url{https://arxiv.org/abs/2507.11704}.
\newblock \bibinfo{note}{(To appear in TPLP 2025)}.

\bibitemdeclare{article}{falilusc20a}
\bibitem{falilusc20a}
\bibinfo{author}{Jorge \surnamestart Fandinno\surnameend},
  \bibinfo{author}{Vladimir \surnamestart Lifschitz\surnameend},
  \bibinfo{author}{Patrick \surnamestart L{\"u}hne\surnameend} \&
  \bibinfo{author}{Torsten \surnamestart Schaub\surnameend}
  (\bibinfo{year}{2020}): \emph{\bibinfo{title}{Verifying Tight Logic Programs
  with anthem and {V}ampire}}.
\newblock {\slshape \bibinfo{journal}{Theory and Practice of Logic
  Programming}} \bibinfo{volume}{20}(\bibinfo{number}{5}), pp.
  \bibinfo{pages}{735--750}, \doi{10.1017/S1471068420000344}.

\bibitemdeclare{book}{bo:Fitting83}
\bibitem{bo:Fitting83}
\bibinfo{author}{Melvin \surnamestart Fitting\surnameend}
  (\bibinfo{year}{1983}): \emph{\bibinfo{title}{Proof Methods for Modal and
  Intuitionistic Logics}}.
\newblock \bibinfo{publisher}{D.~Reidel}, \bibinfo{address}{Dordrecht},
  \doi{10.1007/978-94-017-2794-5}.

\bibitemdeclare{inproceedings}{gellif90a}
\bibitem{gellif90a}
\bibinfo{author}{Michael \surnamestart Gelfond\surnameend} \&
  \bibinfo{author}{Vladimir \surnamestart Lifschitz\surnameend}
  (\bibinfo{year}{1990}): \emph{\bibinfo{title}{Logic Programs with Classical
  Negation}}.
\newblock In \bibinfo{editor}{David \surnamestart Warren\surnameend} \&
  \bibinfo{editor}{P{\'e}ter \surnamestart Szeredi\surnameend}, editors:
  {\slshape \bibinfo{booktitle}{Proceedings of the Seventh International
  Conference on Logic Programming (ICLP'90)}}, \bibinfo{publisher}{{MIT}
  Press}, pp. \bibinfo{pages}{579--597}.

\bibitemdeclare{article}{ar:Gentzen35}
\bibitem{ar:Gentzen35}
\bibinfo{author}{Gerhard \surnamestart Gentzen\surnameend}
  (\bibinfo{year}{1935}): \emph{\bibinfo{title}{Untersuchungen {\"u}ber das
  {L}ogische {S}chlie{\ss}en}}.
\newblock {\slshape \bibinfo{journal}{Mathematische Zeit\-schrift}}
  \bibinfo{volume}{39}, pp. \bibinfo{pages}{176--210, 405--431},
  \doi{10.1007/BF01201353}.

\bibitemdeclare{article}{goedel32a}
\bibitem{goedel32a}
\bibinfo{author}{Kurt \surnamestart G{\"o}del\surnameend}
  (\bibinfo{year}{1932}): \emph{\bibinfo{title}{Zum intuitionistischen
  {A}ussagenkalk{\"u}l}}.
\newblock {\slshape \bibinfo{journal}{Anzeiger der Akademie der Wissenschaften
  in Wien}}, pp. \bibinfo{pages}{65--66}.

\bibitemdeclare{incollection}{heyting30a}
\bibitem{heyting30a}
\bibinfo{author}{Arend \surnamestart Heyting\surnameend}
  (\bibinfo{year}{1930}): \emph{\bibinfo{title}{Die formalen {R}egeln der
  intuitionistischen {L}ogik}}.
\newblock In: {\slshape \bibinfo{booktitle}{Sitzungsberichte der Preussischen
  Akademie der Wissenschaften}}, \bibinfo{publisher}{Deutsche Akademie der
  Wissenschaften zu Berlin}, pp. \bibinfo{pages}{42--56}.

\bibitemdeclare{proceedings}{ijcai16}
\bibitem{ijcai16}
\bibinfo{editor}{Subbarao \surnamestart Kambhampati\surnameend}, editor
  (\bibinfo{year}{2016}): \emph{\bibinfo{title}{Proceedings of the Twenty-fifth
  International Joint Conference on Artificial Intelligence (IJCAI'16)}}.
  \bibinfo{publisher}{IJCAI/AAAI Press}.

\bibitemdeclare{book}{lifschitz19a}
\bibitem{lifschitz19a}
\bibinfo{author}{Vladimir \surnamestart Lifschitz\surnameend}
  (\bibinfo{year}{2019}): \emph{\bibinfo{title}{Answer Set Programming}}.
\newblock \bibinfo{publisher}{Springer}, \bibinfo{address}{Cham},
  \doi{10.1007/978-3-030-24658-7}.

\bibitemdeclare{article}{lipeva01a}
\bibitem{lipeva01a}
\bibinfo{author}{Vladimir \surnamestart Lifschitz\surnameend},
  \bibinfo{author}{David \surnamestart Pearce\surnameend} \&
  \bibinfo{author}{Agust{\'{\i}}n \surnamestart Valverde\surnameend}
  (\bibinfo{year}{2001}): \emph{\bibinfo{title}{Strongly equivalent logic
  programs}}.
\newblock {\slshape \bibinfo{journal}{ACM Transactions on Computational Logic}}
  \bibinfo{volume}{2}(\bibinfo{number}{4}), pp. \bibinfo{pages}{526--541},
  \doi{10.1145/383779.383783}.

\bibitemdeclare{inproceedings}{lipeva07a}
\bibitem{lipeva07a}
\bibinfo{author}{Vladimir \surnamestart Lifschitz\surnameend},
  \bibinfo{author}{David \surnamestart Pearce\surnameend} \&
  \bibinfo{author}{Agust{\'{\i}}n \surnamestart Valverde\surnameend}
  (\bibinfo{year}{2007}): \emph{\bibinfo{title}{A Characterization of Strong
  Equivalence for Logic Programs with Variables}}.
\newblock In \bibinfo{editor}{Chitta \surnamestart Baral\surnameend},
  \bibinfo{editor}{Gerhard \surnamestart Brewka\surnameend} \&
  \bibinfo{editor}{John \surnamestart Schlipf\surnameend}, editors: {\slshape
  \bibinfo{booktitle}{Proceedings of the Ninth International Conference on
  Logic Programming and Nonmonotonic Reasoning (LPNMR'07)}}, {\slshape
  \bibinfo{series}{Lecture Notes in Artificial Intelligence}}
  \bibinfo{volume}{4483}, \bibinfo{publisher}{Springer},
  \bibinfo{address}{Heidelberg}, pp. \bibinfo{pages}{188--200},
  \doi{10.1007/978-3-540-72200-7_17}.

\bibitemdeclare{article}{ar:Mints10}
\bibitem{ar:Mints10}
\bibinfo{author}{Grigori \surnamestart Mints\surnameend}
  (\bibinfo{year}{2010}): \emph{\bibinfo{title}{Cut-free formulations for a
  quantified logic of here and there}}.
\newblock {\slshape \bibinfo{journal}{Ann. Pure Appl. Log.}}
  \bibinfo{volume}{162}(\bibinfo{number}{3}), pp. \bibinfo{pages}{237--242},
  \doi{10.1016/J.APAL.2010.09.009}.

\bibitemdeclare{inproceedings}{inp:Otten97}
\bibitem{inp:Otten97}
\bibinfo{author}{Jens \surnamestart Otten\surnameend} (\bibinfo{year}{1997}):
  \emph{\bibinfo{title}{{\ileantap}: An Intuitionistic Theorem Prover}}.
\newblock In \bibinfo{editor}{Didier \surnamestart Galmiche\surnameend},
  editor: {\slshape \bibinfo{booktitle}{TABLEAUX 1997}}, {\slshape
  \bibinfo{series}{Lecture Notes in Artificial Intelligence}}
  \bibinfo{volume}{1227}, \bibinfo{publisher}{Springer},
  \bibinfo{address}{Heidelberg}, pp. \bibinfo{pages}{307--312},
  \doi{10.1007/BFb0027422}.

\bibitemdeclare{article}{ar:Otten10}
\bibitem{ar:Otten10}
\bibinfo{author}{Jens \surnamestart Otten\surnameend} (\bibinfo{year}{2010}):
  \emph{\bibinfo{title}{Restricting backtracking in connection calculi}}.
\newblock {\slshape \bibinfo{journal}{AI Commun.}}
  \bibinfo{volume}{23}(\bibinfo{number}{2--3}), pp. \bibinfo{pages}{159--182},
  \doi{10.3233/AIC-2010-0464}.

\bibitemdeclare{inproceedings}{inp:Otten11}
\bibitem{inp:Otten11}
\bibinfo{author}{Jens \surnamestart Otten\surnameend} (\bibinfo{year}{2011}):
  \emph{\bibinfo{title}{A Non-clausal Connection Calculus}}.
\newblock In \bibinfo{editor}{Kai \surnamestart Br{\"u}nnler\surnameend} \&
  \bibinfo{editor}{George \surnamestart Metcalfe\surnameend}, editors:
  {\slshape \bibinfo{booktitle}{TABLEAUX 2011}}, {\slshape
  \bibinfo{series}{Lecture Notes in Artificial Intelligence}}
  \bibinfo{volume}{6793}, \bibinfo{publisher}{Springer},
  \bibinfo{address}{Heidelberg}, pp. \bibinfo{pages}{226--241},
  \doi{10.1007/978-3-642-22119-4_18}.

\bibitemdeclare{inproceedings}{inp:Otten16}
\bibitem{inp:Otten16}
\bibinfo{author}{Jens \surnamestart Otten\surnameend} (\bibinfo{year}{2016}):
  \emph{\bibinfo{title}{{{\sf nanoCoP}}: {A} Non-clausal Connection Prover}}.
\newblock In \bibinfo{editor}{Nicola \surnamestart Olivetti\surnameend} \&
  \bibinfo{editor}{Ashish \surnamestart Tiwari\surnameend}, editors: {\slshape
  \bibinfo{booktitle}{IJCAR 2016}}, {\slshape \bibinfo{series}{Lecture Notes in
  Artificial Intelligence}} \bibinfo{volume}{9706},
  \bibinfo{publisher}{Springer}, \bibinfo{address}{Heidelberg}, pp.
  \bibinfo{pages}{302--312}, \doi{10.1007/978-3-319-40229-1_21}.

\bibitemdeclare{inproceedings}{inp:Otten17b}
\bibitem{inp:Otten17b}
\bibinfo{author}{Jens \surnamestart Otten\surnameend} (\bibinfo{year}{2017}):
  \emph{\bibinfo{title}{Non-clausal Connection Calculi for Non-classical
  Logics}}.
\newblock In \bibinfo{editor}{Renate \surnamestart Schmidt\surnameend} \&
  \bibinfo{editor}{Cl\'{a}udia \surnamestart Nalon\surnameend}, editors:
  {\slshape \bibinfo{booktitle}{TABLEAUX 2017}}, {\slshape
  \bibinfo{series}{Lecture Notes in Artificial Intelligence}}
  \bibinfo{volume}{10501}, \bibinfo{publisher}{Springer},
  \bibinfo{address}{Cham}, pp. \bibinfo{pages}{209--227},
  \doi{10.1007/978-3-319-66902-1_13}.

\bibitemdeclare{inproceedings}{inp:Otten21}
\bibitem{inp:Otten21}
\bibinfo{author}{Jens \surnamestart Otten\surnameend} (\bibinfo{year}{2021}):
  \emph{\bibinfo{title}{The {{\sf nanoCoP} 2.0} Connection Provers for
  Classical, Intuitionistic and Modal Logics}}.
\newblock In \bibinfo{editor}{Anupam \surnamestart Das\surnameend} \&
  \bibinfo{editor}{Sara \surnamestart Negri\surnameend}, editors: {\slshape
  \bibinfo{booktitle}{TABLEAUX 2021}}, {\slshape \bibinfo{series}{Lecture Notes
  in Artificial Intelligence}} \bibinfo{volume}{12842},
  \bibinfo{publisher}{Springer}, \bibinfo{address}{Cham}, pp.
  \bibinfo{pages}{236--249}, \doi{10.1007/978-3-030-86059-2\_14}.

\bibitemdeclare{article}{ar:OttenBibel03}
\bibitem{ar:OttenBibel03}
\bibinfo{author}{Jens \surnamestart Otten\surnameend} \&
  \bibinfo{author}{Wolfgang \surnamestart Bibel\surnameend}
  (\bibinfo{year}{2003}): \emph{\bibinfo{title}{{{\sf leanCoP}}: lean
  connection-based theorem proving}}.
\newblock {\slshape \bibinfo{journal}{Journal of Symbolic Computation}}
  \bibinfo{volume}{36}(\bibinfo{number}{1--2}), pp. \bibinfo{pages}{139--161},
  \doi{10.1016/S0747-7171(03)00037-3}.

\bibitemdeclare{inproceedings}{inp:OttenBibel17}
\bibitem{inp:OttenBibel17}
\bibinfo{author}{Jens \surnamestart Otten\surnameend} \&
  \bibinfo{author}{Wolfgang \surnamestart Bibel\surnameend}
  (\bibinfo{year}{2017}): \emph{\bibinfo{title}{Advances in Connection-Based
  Automated Theorem Proving}}.
\newblock In \bibinfo{editor}{Mike \surnamestart Hinchey\surnameend},
  \bibinfo{editor}{Jonathan~P. \surnamestart Bowen\surnameend} \&
  \bibinfo{editor}{Ernst-R{\"u}diger \surnamestart Olderog\surnameend},
  editors: {\slshape \bibinfo{booktitle}{Provably Correct Systems}},
  \bibinfo{series}{NASA Monographs in Systems and Software Engineering},
  \bibinfo{publisher}{Springer}, \bibinfo{address}{Cham}, pp.
  \bibinfo{pages}{211--241}, \doi{10.1007/978-3-319-48628-4_9}.

\bibitemdeclare{article}{pearce06a}
\bibitem{pearce06a}
\bibinfo{author}{David \surnamestart Pearce\surnameend} (\bibinfo{year}{2006}):
  \emph{\bibinfo{title}{Equilibrium logic}}.
\newblock {\slshape \bibinfo{journal}{Annals of Mathematics and Artificial
  Intelligence}} \bibinfo{volume}{47}(\bibinfo{number}{1--2}), pp.
  \bibinfo{pages}{3--41}, \doi{10.1007/s10472-006-9028-z}.

\bibitemdeclare{inproceedings}{peguva00a}
\bibitem{peguva00a}
\bibinfo{author}{David \surnamestart Pearce\surnameend},
  \bibinfo{author}{Inmaculada \surnamestart {de Guzm{\'a}n}\surnameend} \&
  \bibinfo{author}{August{\'i}n \surnamestart Valverde\surnameend}
  (\bibinfo{year}{2000}): \emph{\bibinfo{title}{A Tableau Calculus for
  Equilibrium Entailment}}.
\newblock In \bibinfo{editor}{Roy \surnamestart Dyckhoff\surnameend}, editor:
  {\slshape \bibinfo{booktitle}{Proceedings of the Ninth International
  Conference on Automated Reasoning with Analytic Tableaux and Related Methods
  (TABLEAUX 2000)}}, {\slshape \bibinfo{series}{Lecture Notes in Computer
  Science}} \bibinfo{volume}{1847}, \bibinfo{publisher}{Springer},
  \bibinfo{address}{Heidelberg}, pp. \bibinfo{pages}{352--367},
  \doi{10.1007/10722086_28}.

\bibitemdeclare{article}{peaval05a}
\bibitem{peaval05a}
\bibinfo{author}{David \surnamestart Pearce\surnameend} \&
  \bibinfo{author}{August{\'i}n \surnamestart Valverde\surnameend}
  (\bibinfo{year}{2005}): \emph{\bibinfo{title}{A First Order Nonmonotonic
  Extension of Constructive Logic}}.
\newblock {\slshape \bibinfo{journal}{Studia Logica}}
  \bibinfo{volume}{30}(\bibinfo{number}{2--3}), pp. \bibinfo{pages}{321--346},
  \doi{10.1007/s11225-005-8473-8}.

\bibitemdeclare{article}{ar:Pelletier86}
\bibitem{ar:Pelletier86}
\bibinfo{author}{Francis~Jeffry \surnamestart Pelletier\surnameend}
  (\bibinfo{year}{1986}): \emph{\bibinfo{title}{Seventy-Five Problems for
  Testing Automatic Theorem Provers}}.
\newblock {\slshape \bibinfo{journal}{Journal of Automated Reasoning}}
  \bibinfo{volume}{2}(\bibinfo{number}{2}), pp. \bibinfo{pages}{191--216},
  \doi{10.1007/BF02432151}.

\bibitemdeclare{article}{ar:RathsOttenKreitz07}
\bibitem{ar:RathsOttenKreitz07}
\bibinfo{author}{Thomas \surnamestart Raths\surnameend}, \bibinfo{author}{Jens
  \surnamestart Otten\surnameend} \& \bibinfo{author}{Christoph \surnamestart
  Kreitz\surnameend} (\bibinfo{year}{2007}): \emph{\bibinfo{title}{The {ILTP}
  problem library for intuitionistic logic}}.
\newblock {\slshape \bibinfo{journal}{Journal of Automated Reasoning}}
  \bibinfo{volume}{38}, pp. \bibinfo{pages}{261--271},
  \doi{10.1007/s10817-006-9060-z}.

\bibitemdeclare{book}{bo:Smullyan68}
\bibitem{bo:Smullyan68}
\bibinfo{author}{Raymond~M. \surnamestart Smullyan\surnameend}
  (\bibinfo{year}{1968}): \emph{\bibinfo{title}{First-Order Logic}}.
\newblock \bibinfo{series}{Ergebnisse der Mathematik und ihrer Grenzgebiete},
  \bibinfo{publisher}{Springer}, \bibinfo{address}{Berlin, Heidelberg, New
  York}, \doi{10.1007/978-3-642-86718-7}.

\bibitemdeclare{book}{bo:Wallen90}
\bibitem{bo:Wallen90}
\bibinfo{author}{Lincoln~A. \surnamestart Wallen\surnameend}
  (\bibinfo{year}{1990}): \emph{\bibinfo{title}{Automated Deduction in
  Nonclassical Logics}}.
\newblock \bibinfo{publisher}{MIT Press}, \bibinfo{address}{Cambridge, Mass.}

\end{thebibliography}


\end{document}